# Experimental signature of layer skyrmions and implications for band topology in twisted $WSe_2$ bilayers

Fan Zhang[1†], Nicolás Morales-Durán[1†], Yanxing Li[1†], Wang Yao[2], Jung-Jung Su[3], Yu-Chuan Lin[4,5], Chengye Dong[4], Xiaohui Liu[1], Fu-Xiang Rikudo Chen[1], Hyunsue Kim[1], Kenji Watanabe[6], Takashi Taniguchi[6], Xiaoqin Li[1], Joshua A. Robinson[4], Allan H. Macdonald[1*], and Chih-Kang Shih[1*]

[1]Department of Physics, University of Texas at Austin, Austin, TX, USA

[2]Department of Physics, The University of Hong Kong, Hong Kong, China

[3]Department of Electrophysics, National Yang Ming Chiao Tung University, Hsinchu, Taiwan

[4]Department of Materials Science and Engineering, Pennsylvania State University, University Park, PA, USA

[5]Department of Materials Science and Engineering, National Yang Ming Chiao Tung University, Hsinchu, Taiwan

[6]Research Center for Materials Nanoarchitectonics, National Institute for Materials Science, Tsukuba, Japan

[†]These authors contribute equally to this work.

*Corresponding authors' email: macdpc@physics.utexas.edu; shih@physics.utexas.edu

## Abstract

Transition metal dichalcogenide (TMD) twisted homobilayers have been established as an ideal platform for studying strong correlation phenomena, as exemplified by the recent discovery of fractional Chern insulator (FCI) states in twisted $MoTe_2$[1–4] and Chern insulators (CI)[5] and unconventional superconductivity[6,7] in twisted $WSe_2$ ($tWSe_2$). In these systems, nontrivial topology in the strongly layer-hybridized regime can arise from a spatial patterning of interlayer tunneling amplitudes and layer-dependent potentials that yields a lattice of layer skyrmions. Here we report on experimental signatures of skyrmion textures in the layer degree of freedom of Rhombohedral-stacked (R-stacked) $tWSe_2$ homobilayers. This observation is based on scanning tunneling spectroscopy that separately resolves the Γ-valley and K-valley moiré electronic states. We show that Γ-valley states are subjected to a moiré potential with an amplitude of ~ 120 meV. At ~150 meV above the Γ-valley, the K-valley states are subjected to a weaker moiré potential of ~30 meV. Most significantly, we reveal opposite layer polarizations of the K-valley at the MX and XM sites within the moiré unit cell, confirming the theoretically predicted layer skyrmion texture. The dI/dV mappings allow the parameters that enter the continuum model of moiré bands in twisted TMD bilayers to be determined experimentally, further establishing a direct correlation between the shape of the LDOS profile in real space and the topology of topmost moiré band.

## Main

Semiconducting transition metal dichalcogenide (TMD) moiré materials have attracted significant attention within the last few years, as they offer a tunable platform to study quantum phenomena arising from correlation and topology[8–15]. In particular, some R-stacked twisted homobilayers in which the top of the spin-valley locked valence band comes from the K (−K)-valley, like $MoTe_2$ and $WSe_2$, provide an opportunity to explore the realm of topologically-ordered states[13,16–31], as experimentally confirmed with the observation of the fractional quantum anomalous Hall effect (FQAHE)[2–4] and the fractional quantum spin Hall effect (FQSHE)[32]. Continuum-model theoretical descriptions of R-stacked homobilayers include a local layer-dependent term that contains one layer skyrmion per moiré unit cell[10,33,34]. Subsequent studies have shown that this skyrmion structure plays a crucial role in determining the topological character of the band[10,33–35], giving rise to a real-space Berry curvature of quantized flux per moiré unit cell. Because it is able to probe both atomic structure and local electronic properties, scanning tunneling microscopy (STM) is an ideal tool to investigate this unique skyrmion-like layer pseudospin. However, success in executing this program requires the ability to separate Γ-valley and K-valley state contributions, and this is non-trivial due to the valley dependent tunneling decay coefficient.

In this work we perform STM/S measurements to directly probe the layer polarizations at MX and XM positions, thereby revealing the layer-skyrmion texture in R-stacked $WSe_2$ bilayers twisted to 5.1 and 4.2 degrees, in agreement with theoretical modeling. Following previously developed methodology[36,37], Γ- and K-valley states are resolved individually by combining conventional STS, which is more sensitive to the Γ-valley states, with constant current STS, which greatly enhances the sensitivity to the K-valley states. This capability allows us to reveal the moiré potential modulations for Γ-valley and K-valley separately. In addition, the layer skyrmion texture is manifested by opposite polarizations of the K-valley wavefunction between the MX and XM sites. This direct experimental observation of the pseudospin texture and the moiré potential modulation sheds light on the parameters of the continuum model theoretical description[10,38], which is key to experimental searches for exotic topological phases (e.g. FQAHE) in twisted TMD systems. This unique feedback loop may deepen the understanding of the conditions required for the emergence of Chern bands and promote the discovery of exotic new topological phases in moiré TMDs.

**Theoretical modeling of twisted TMD homobilayers**

We first discuss the continuum model description of the K-valleys in twisted TMD homobilayers, from which the layer pseudospin picture emerges. At a small twist angle $\theta$, a moiré superlattice with a lattice constant $a_M \sim 10$ nm will be created as shown in Fig. 1a. Top view and side view of the bilayer structure for three high symmetry configurations (marked as AA, MX, XM) are shown in Fig. 1b. At the MX site, the transition metal (M) of the top layer is aligned with the chalcogen atom (X) of the bottom layer. While at the XM site, the chalcogen atom (X) of the top layer is aligned with the transition metal (M) of the bottom layer. The AA site corresponds to the stacking at which the transition metal atoms from both layers are aligned. Figure 1c shows the moiré Brillouin zone (MBZ) emerging from the relative twist between the top and bottom layer BZ. The K points from the top layer ($K_t$) and bottom layer ($K_b$) are located at adjacent MBZ corners, as illustrated in Fig. 1c. Due to spin-valley locking[39], the topmost valence bands in K and −K valleys have opposite spin orientations (up for K, down for −K). Because the K and −K valleys are related by time-reversal symmetry, we can focus on the Hamiltonian for the K valley, written in layer space as

$$H_K = \begin{pmatrix} -\frac{\hbar^2(\boldsymbol{k}-\boldsymbol{K}_b)^2}{2m^*} + \Delta_b(\boldsymbol{r}) & \Delta_T(\boldsymbol{r}) \\ \Delta_T^\dagger(\boldsymbol{r}) & -\frac{\hbar^2(\boldsymbol{k}-\boldsymbol{K}_t)^2}{2m^*} + \Delta_t(\boldsymbol{r}) \end{pmatrix}. \tag{1}$$

After applying a unitary transformation, $U_0 = diag(e^{i\boldsymbol{K}_b\cdot \boldsymbol{r}},\ e^{i\boldsymbol{K}_t\cdot \boldsymbol{r}})$, to remove the momentum shifts in the diagonal, the Hamiltonian is given by

$$H_K = -\frac{\hbar^2\boldsymbol{k}^2}{2m^*}\sigma_0 + \boldsymbol{\Delta}(\boldsymbol{r})\cdot\sigma + \Delta_0\sigma_0, \tag{2}$$

where $\sigma = (\sigma_x, \sigma_y, \sigma_z)$ are the layer Pauli matrices and $\sigma_0$ is the identity. The layer-dependent moiré potentials and the interlayer tunneling are respectively given by

$$\Delta_{b/t}(\boldsymbol{r}) = \sum_{j=1,3,5} 2V_m \cos(\boldsymbol{b}_j \cdot \boldsymbol{r} \pm \psi), \tag{3}$$

$$\Delta_T(\boldsymbol{r}) = \omega\left(e^{i\boldsymbol{q}_1\cdot\boldsymbol{r}} + e^{i\boldsymbol{q}_2\cdot\boldsymbol{r}} + e^{i\boldsymbol{q}_3\cdot\boldsymbol{r}}\right), \tag{4}$$

With $\boldsymbol{q}_1 = \boldsymbol{K}_b - \boldsymbol{K}_t$ , $\boldsymbol{q}_2 = \boldsymbol{b}_2 + \boldsymbol{q}_1$, $\boldsymbol{q}_3 = \boldsymbol{b}_3 + \boldsymbol{q}_1$, where the $\boldsymbol{b}_j$ belong to the first shell of reciprocal lattice vectors. We have defined $\boldsymbol{\Delta}(\boldsymbol{r}) = (Re\Delta_T, Im\Delta_T, (\Delta_b - \Delta_t)/2)$ and $\Delta_0 = (\Delta_b + \Delta_t)/2$. When the layer degree of freedom is understood as a pseudospin, the continuum model Hamiltonian corresponds to a fermion in the presence of an effective Zeeman field $\boldsymbol{\Delta}$ acting on the layer pseudospin, whose spatial structure is shown in Fig. 1d. This layer-Zeeman field forms a texture with one skyrmion per moiré unit cell, with its bottom component vanishing at the XM site and its top component vanishing at the MX site. If the layer-pseudospin of holes aligns with the direction of $\boldsymbol{\Delta}$, the corresponding wave functions will acquire a winding as they cover the moiré unit cell, resulting in a real space Berry curvature that plays the role of an effective magnetic field with quantized flux per moiré cell[28].

It has been suggested that the particular spatial form of the layer pseudospin in TMD homobilayers is crucial for the emergence of non-zero Chern numbers in the moiré mini bands. This new perspective to understand the physics of TMD homobilayers could provide new insights into the nature of the topologically-ordered states that have been observed in these systems. However, up to this point, there is no direct experimental evidence to confirm that the layer pseudospin state covers its full Bloch sphere, with opposite full layer polarizations (layer-pseudospins) at the XM and MX sites, arising from alignment between layer pseudospins and layer-skyrmion fields.

The presence of the skyrmion texture in the layer-pseudospin degree of freedom does not guarantee the emergence of topological bands. The tunneling and moiré potentials in the continuum model depend on three material-specific parameters $(V_m, \psi, \omega)$, ($V_m$: moiré potential amplitude, $\psi$: moiré potential phase, $\omega$: interlayer hopping) and the Chern number of the topmost moiré mini band depends crucially on the interplay of these parameters. The usual approach to estimate the continuum model parameters is via large-scale ab initio electronic structure calculations that have all the uncertainties of DFT theory. Here we show that local probe measurements provide a route to extract values directly from the experiment.

**STM observation of layer-pseudospin skyrmions**

We now discuss our experimental observation of layer-pseudospin textures. Figure 2a shows the schematic STM experimental measurement set-up of a small angle $tWSe_2$ homobilayer. The samples were either synthesized through metal organic chemical vapor deposition (MOCVD) on an epitaxial graphene substrate (See in Supplementary Information, Fig. S1 and S2) or fabricated through exfoliation-and-stacking method (Supplementary Fig. S3-5). Shown in Fig. 2b is an STM image of a moiré superlattice with a wavelength $a_M = 3.71nm$. From the measured value of $a_M$ the twist angle $\theta = 5.1°$ can be extracted through $a_M = a/(2\,sin\left(\frac{\theta}{2}\right))$, where the $WSe_2$ monolayer lattice constant $a$ is $0.328nm$. Shown in Fig. 2c are tunneling spectra acquired at four sites marked as AA, MX, Bridge (Br), XM in Fig. 2b. Here the spectra are acquired using the conventional STS mode where the tip-to-sample distance is maintained at a constant value while the sample bias is varied (referred to as a constant height STS or CH-STS). As shown in Fig. 2c, the energy of the topmost Γ band is lower at the AA sites than that at the MX/XM sites by an amount of ~120 meV, corresponding to a moiré potential modulation amplitude of 120 meV. This results from the larger interlayer spacing at the AA sites, which lowers the hybridization strength. This value for the Γ− moiré modulation strength agrees well with recent reports[40,41]. Electronic state modulation near the K points has not been previously reported because states near the K-point decay very rapidly into the vacuum as discussed by Zhang et al[36]. A direct measurement of the decay constant for Γ- and K-valley states is shown in Fig. 2d. At a typical tip-to-sample distance of 10 Å, the tunneling probability of K-valley states is several orders of magnitude smaller than the Γ-valley states. This low sensitivity can be overcome by taking STS in the constant current mode (referred to as CC-STS) which automatically brings the tip-to-sample distance closer till the states in the K-valley are available for tunneling. The variation of the tip-to-sample-distance (Z) as a function of the sample bias is shown in Fig. 2e, where a transition from Γ- to K-valley tunneling is indicated by a rapid reduction in Z by ~ 3Å (vertical red dashed line). Upon passing the K-valley, another transition takes place as the tunneling occurs between the tip and the underlying graphene. Additionally, the differential conductivity spectra, $(\partial I/\partial V)_I$, shown in Fig.

2f, reveals a K valley modulation of around 30 meV—much smaller than for the Γ-valley states—due to weaker interlayer hybridization at the K-valley. At the K valley the spectrum acquired at the XM site is of higher intensity than that at the MX site, due to layer-polarized K-valley wavefunctions (layer pseudospin) resulting from a combination of weak interlayer hybridization and the effective electric field arising from interlayer charge transfer[42,43]. Significantly, the AA site has the highest spectral intensity. Such a variation of spectral intensity is also consistent with the Z-V behavior at the K-Gr (graphene) transition (marked by the orange arrow for site AA, which shows largest $\Delta Z$ change. This point is further discussed in Supplementary Figures S6 and S7.) Measurements carried out on exfoliated/stacked sample with $\theta_t = 4.2°$ degree show essentially the same behavior except for a longer moiré wavelength of 4.45 nm, as depicted in Figure S5.

Shown in Fig. 3a is the colormap of CH-STS spectra along high-symmetry directions of the moiré superlattice [AA-MX-Br-XM-AA]. The energy difference of the topmost Γ bands, between AA site and MX/XM sites reflects the large moiré potential modulation amplitude ~0.12 eV of the Γ valley. The color map of CC-STS shows that the K-valley states are located about 0.15 eV above the Γ valley, with a weaker (~ 30 meV) variation in peak position (Fig. 3b). Crucially, the intensity exhibits a striking contrast, especially between MX and XM sites, reflecting opposite layer polarizations with a dipole pointing down at the MX site and up at the XM site, resulting in opposite wavefunction localization at the top (XM) and bottom (MX) layers.

We next investigate the 2D texture of the K valley (2D texture of the Γ valley is discussed in Supplementary Fig. S8). Shown in Fig. 3c is the STM topography acquired at -0.85V. This bias corresponds to the black dashed line shown in Fig. 3b where the tunneling primarily results from electronic states at the K valley. Shown in Fig. 3d is the conductivity image acquired at -0.85 V, at a constant height, and with a lock-in modulation amplitude of 20 mV. The modulation bandwidth is large enough to cover most of the states near the VBM and the image can be viewed as capturing the spatial modulation of the local density of states near the VBM. Acquiring the conductivity image at a constant height also removes the uncertainty associated with a varying tip-to-sample distance in the topography image (Also see the constant current conductivity image in Supplementary Fig. S9). In this conductivity image atomic lattice corrugations are also observed which can be removed by filtering out the Bragg peak in the FFT (shown as Fig. 3e). Note that the intensity difference between the XM and MX sites reflects a layer polarization difference and can be used to represent the texture of the layer pseudospins. (The same argument does not apply to the K-valley states at other sites (AA and Br) since the local DOS also varies with position.) Regardless, the contrast between the MX and XM sites reflects the skyrmion texture of the layer pseudospin degree of freedom. Note that the z-component is a physical observable corresponding to electrical polarization (therefore measurable using STS). The azimuthal component, which corresponds to the phase difference between electrons in different layers, is not measurable by STS. Further details are discussed in Supplementary Note 1.

To better capture the pseudospin texture, we interpret the STM results by comparing with theoretical simulations using parameterized continuum models.

**Continuum model parameters from STS**

We interpret the dI/dV map shown Fig. 3e as a layer-projected LDOS and obtained the dI/dV line profile shown in Fig. 4a along AA-XM-Br-MX-AA by averaging different lines across the sample to smoothen out possible inhomogeneities. We compare this spatial profile, also shown as the black solid curve in Fig. 4b, with the local maximum of the theoretical LDOS near $K_{VBM}$

$$D(\boldsymbol{r},\varepsilon) = \mathcal{N}\sum_{\boldsymbol{k}} |\psi_{\boldsymbol{k}}^{t}(\boldsymbol{r})|^2 exp\left(-\frac{(\varepsilon-\varepsilon_{\boldsymbol{k}})^2}{2\sigma^2}\right), \tag{5}$$

calculated using the continuum model and adjusting parameters to match experiment. Here $\psi_{\boldsymbol{k}}^{t}(\boldsymbol{r})$ is the top-layer component of the continuum model eigenstate, $\varepsilon_{\boldsymbol{k}}$ is a continuum model energy and $\mathcal{N}$ is a normalization constant. Here we have assumed a Gaussian resolution window for our scanning probe tip with a width $\sigma$ of around 20 meV. The best theoretical fit is plotted as a red dashed line in Fig. 4b. From this fit we estimate the continuum model parameters for 5.08° $tWSe_2$ to be $V_m$ = 13.6 meV, $\psi = 49.1°$ degrees $\omega$ = 10.0 meV. (The fitting procedure is explained in Supplementary Fig. S10. The effective mass of the carriers is also fit and we find $m = 0.6m_0$) Figure 4c shows the moiré band structure corresponding to these parameters, which displays a topmost mini band with a ~50 meV bandwidth. The Chern number of the topmost band is zero, indicating a topologically trivial band. These parameters lead to the pseudospin texture shown in Fig. 4d. The false color rendition represents the z-component of the layer pseudospin while the in-plane component is labeled by the black arrows. (This fitting procedure can also be applied to the 4.2-degree twisted $WSe_2$ and the results are shown in Supplementary Fig. S11.) By this route, scanning tunneling microscopy techniques can be used to reveal the topological character of moiré TMD samples. The sensitivity of the continuum model fit to the experimental data is illustrated in Figures 4e-h, where we plot the layer-projected LDOS as a function of position, for different values of the phase parameter $\psi$. Figure 4e shows the Chern number of the topmost moiré band from the continuum model as a function of $\psi$ and $V_m/\omega$ at $\theta$=5.08°. The shape of the LDOS profile reflects the topology of the band. In the trivial regime (C = 0) the LDOS maximum is consistently located at AA site, whereas in the topological regime (C = +1 or -1) the LDOS maximum shifts to the XM or MX sites and this is reflected in our fit by a value of $\psi$ that is smaller than some DFT estimates. Due to variations in the twisted TMD samples—such as twist angle, interlayer coupling strength, and strain—the LDOS at the AA site in the topological regime can either be a minimum (Figure 4g) or lie between the values at the XM and MX sites (Figure 4f). These cases are consistent with recent reports by Thompson et al[44] and Wang et al[45] , which examined small twist angle $MoTe_2$ bilayers known to be Chern insulators[1–4]. These studies found a very similar Γ-valley moiré potential and a comparable pseudospin texture for the K-valley, whose relationship between band topology and pseudospin texture fit in well with our method (Figure 4f-g). In this context, these two independent investigations, along with our findings, corroborate each other well. Additionally, the layer-pseudospin of XM and MX can even flip sign due to the competition between the out-of-plane ferroelectric effect and the strain-induced in-plane piezoelectric effect. In that case, the Chern number also flips sign (Figure 4h). The effect of changing the ratio $V_m/\omega$ on the fitted continuum model parameters is discussed in Fig. S12, where we plot the layer-projected LDOS as a function of position for different sets of parameters, confirming that the phase parameter $\psi$ is the most crucial to the band topology.

## Summary and Discussion

In summary, we performed scanning tunneling microscopy/spectroscopy (STM/STS) to probe the local electronic states of small angle $tWSe_2$, providing evidence for the presence of skyrmions in the layer degree of freedom. By using the local DOS texture measured by STS as the input, we determine the parameters for the continuum model from which the layer pseudospin skyrmion texture as well as the topology of the topmost moiré band can be determined. This combined STS/continuum model approach thus provides a universal framework to unveil the topological nature of the small twist angle TMD bilayers. The applicability of this approach is confirmed not only by the experimental results presented here on $tWSe_2$ bilayers but also by two recent experimental studies of $tMoTe_2$ bilayers[44,45]. In terms of designing strongly correlated topological states e.g. FCI from first principles, our method could be a good alternative to DFT calculations, because DFT calculations are unlikely to be sufficiently accurate due to challenges including non-local exchange interactions. Our method will play a pivotal role in sorting out the interaction physics of many fundamentally interesting states.

We note a recent transport study reported unconventional superconductivity on 5° twisted $WSe_2$[6]. In that study and several following theoretical works nontrivial band topology was assumed, based on continuum model parameters proposed by Devakul et. al.[16]. However, there are no suggestions in the experimental data indicating that the topmost band is topological. In our view, the band topology is not essential for superconductivity to emerge. While a nontrivial topology in superconductivity would be an interesting scenario, our experimental results and reference 6 do not provide any evidence for such topological effects. Another point is that, even with the same twist angle, there could be local Chern number fluctuations (sample variations from substrate, strain, disorder) as reported in magic-angle bilayer graphene[46]. Our measurements reflect the Chern number in the local region, contrasting the averaged result from the transport measurements.

## Methods

### Sample growth for STM

High-quality buffer on SiC was synthesized using a two-step process. First, the monolayer epitaxial graphene was synthesized using silicon sublimation from the Si face of the semi-insulating SiC substrates (II–VI). Before the growth, the SiC substrates were annealed in 10% hydrogen (balance argon) at 1,500 °C for 30 min to remove subsurface damages due to chemical and mechanical polishing. Then monolayer epitaxial graphene (MLEG) was formed at 1,800 °C for 30 min in a pure argon atmosphere. Second, an Ni stressor layer was used to exfoliate the top graphene layer to obtain fresh and high-quality buffer on SiC. After this, 270 nm of Ni was e-beam deposited on MLEG at a rate of 5 Å $s^{-1}$ as a stressor layer. Then a thermal release tape was used to peel off the top graphene layer from the substrate. The growth of $WSe_2$ crystals on an epitaxial graphene substrate was carried out at 800 °C in a custom-built vertical cold-wall chemical vapor deposition (CVD) reactor for 20 min[47]. The tungsten hexacarbonyl ($W(CO)_6$) (99.99%, Sigma-Aldrich) source was kept inside a stainless-steel bubbler in which the temperature and pressure of the bubbler were always held at 37 °C and 730 torr, respectively. Mass-flow controllers were used to supply $H_2$ carrier gas to the bubbler to transport the $W(CO)_6$ precursor into the CVD chamber. The flow rate of the $H_2$ gas through the bubbler was maintained at a constant 8 standard cubic centimeters per minute (sccm), which resulted in a $W(CO)_6$ flow rate of $9.0 \times 10^{-4}$ sccm at the outlet of the bubbler. $H_2Se$ (99.99%, Matheson) gas was supplied from a separate gas manifold and introduced at the inlet of the reactor at a constant flow rate of 30 sccm.

### 4.2-degree t$WSe_2$ sample fabrication

The metal gates consisting of 17 nm thick chromium (Cr) and 11 nm thick platinum (Pt) layers were prepared using photolithography and electron-beam (e-beam) evaporation. Graphite contacts (~10 nm thick) and a hexagonal boron nitride (hBN) bottom layer (~20 nm thick) were first exfoliated onto a 285 nm $SiO_2$/Si substrate and stacked onto the metal gates via the PC dry transfer technique. The substrates were annealed at 400°C and cleaned with an atomic force microscope (AFM) tip to remove residues. Subsequently, a $WSe_2$ monolayer was exfoliated using PDMS onto a 285 nm $SiO_2$/Si substrate. The tear-and-stack method was used to create a twisted $WSe_2$ bilayer with a controlled twist angle with an accuracy better than 0.1° and transferred onto the graphite contact layer using PPC polymer. The PPC was removed by immersion in an acetone solution, followed by and AFM cleaning and annealing at 250°C in an ultrahigh vacuum (about $1 \times 10^{-10}$ torr) chamber.

### STM and STS measurements

STM and STS measurements on 5.1-degree MOCVD grown sample were conducted in a home-built STM at 4.3 K in the ultra-high vacuum chamber, with a base pressure of $2.0x10^{-11}$ torr. STM and STS measurements on 4.2-degree exfoliated/stacked sample were conducted at 77 K in an Omicron LT STM. The W tip was prepared by electrochemical etching and then cleaned by in situ electron-beam heating. STM dI/dV spectra were measured using a standard lock-in technique, for which the modulation frequency was 758 Hz. Two different modes of STS were simultaneously used: (1) the conventional constant-height STS and (2) the constant-current STS.

### Acknowledgement

F.Z., N.M.D, Y.L., F.-X. C., H.K. A.H.M and C.-K.S. were supported by the NSF through the Center for Dynamics and Control of Materials: an NSF Materials Research Science and Engineering Center under cooperative agreement no. DMR-2308817. F.Z., Y.L., F.-X. C., H.K. and C.-K.S. also acknowledge the support from the US Air Force grant no. FA2386-21-1-4061, NSF grant nos. DMR-1808751 and DMR-2219610, and the Welch Foundation F-2164. N.M.D and A.H.M were supported by the U.S. Department of Energy Office of Basic Energy Sciences under Award DE-SC0019481. C. D. and J. A. R. were supported by the Penn State Center for Nanoscale Science (NSF grant no. DMR-2011839) and the Penn State 2DCC-MIP (NSF grant no.

DMR-1539916). Y.-C.L. acknowledges the support from the Center for Emergent Functional Matter Science (CEFMS) of NYCU and the Yushan Young Scholar Program from the Ministry of Education of Taiwan. X.L. and X-Q.L. acknowledge the support from the National Science Foundation through the Center for Dynamics and Control of Materials: an NSF MRSEC under Cooperative Agreement No. DMR – 2308817 for facilitating collaborations between Shih and Li groups. X-Q.L. gratefully acknowledges partial support from the Department of Energy, Office of Basic Energy Sciences under grant DE-SC0019398 for sample fabrication and support for materials supplies from the Welch Foundation chair F-0014.

**Author Contributions**

F.Z., Y.L. and F.-X. C., carried out the STM and STS measurements under the supervision of C.-K.S. N.M.D performed the theoretical modeling under the supervision of A. H. M. Y.-C.L. synthesized the twisted $WSe_2$ bilayers using MOCVD. C.D. prepared the graphitic buffer layer/SiC. J.A.R. supervised the sample preparation. H.K. annealed and pre-treated the sample. X. Liu, Y.L. and F.Z. prepared and annealed the exfoliated/stacked sample. X-Q. Li was involved in the discussion. K.W. and T.T. synthesized the hBN bulk crystals. Y.L., F.Z., and C.-K.S. analyzed the STM data. W. Y. and J.-J. Su involved in the discussion of layer pseudospin. F.Z., N.M.D, Y.L., A. H. M. and C.-K.S. wrote the paper with contributions from all the authors. † These authors contributed equally to this work.

**Competing Interests:** All the authors declare no competing interests.

**Data Availability:** Source data that reproduces the plots in the main text are provided with this paper. All other data that support the findings of this study are available from the corresponding author upon request.

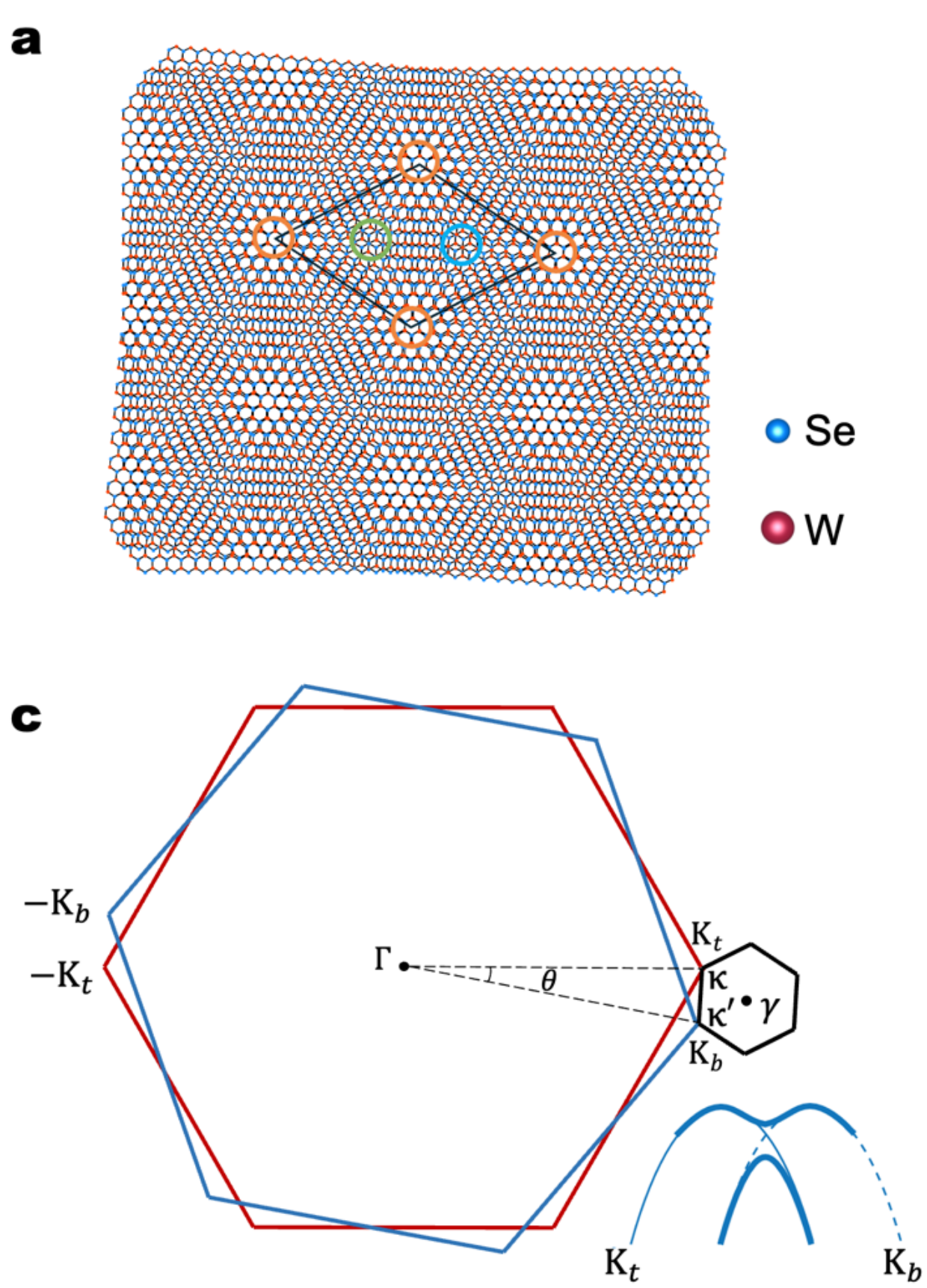

**Figure 1 | Moiré superlattice and layer pseudospin. a,** Schematic of a moiré superlattice in a 5.1-degree twisted $WSe_2$ homobilayer. The high-symmetry AA, MX and XM sites are respectively marked with orange, blue and green circles. **b,** Top and and side views of the structures AA, MX and XM stacking. **c,** Brillouin zones of the top (red) and bottom (blue) layers in a twisted bilayer, and the moiré Brillouin zone (black). Bottom right: schematic of the band hybridization between top and bottom layer K valley valence bands. **d,** Top, Layer pseudospin texture of small angle twisted bilayer $WSe_2$. The arrows layer pseudospin orientations and the color codes its z component. Bottom, corresponding layer distributions of valence band edge carriers (yellow isosurfaces), at the three high symmetry stackings, with the arrows indicating the layer pseudospin orientations.

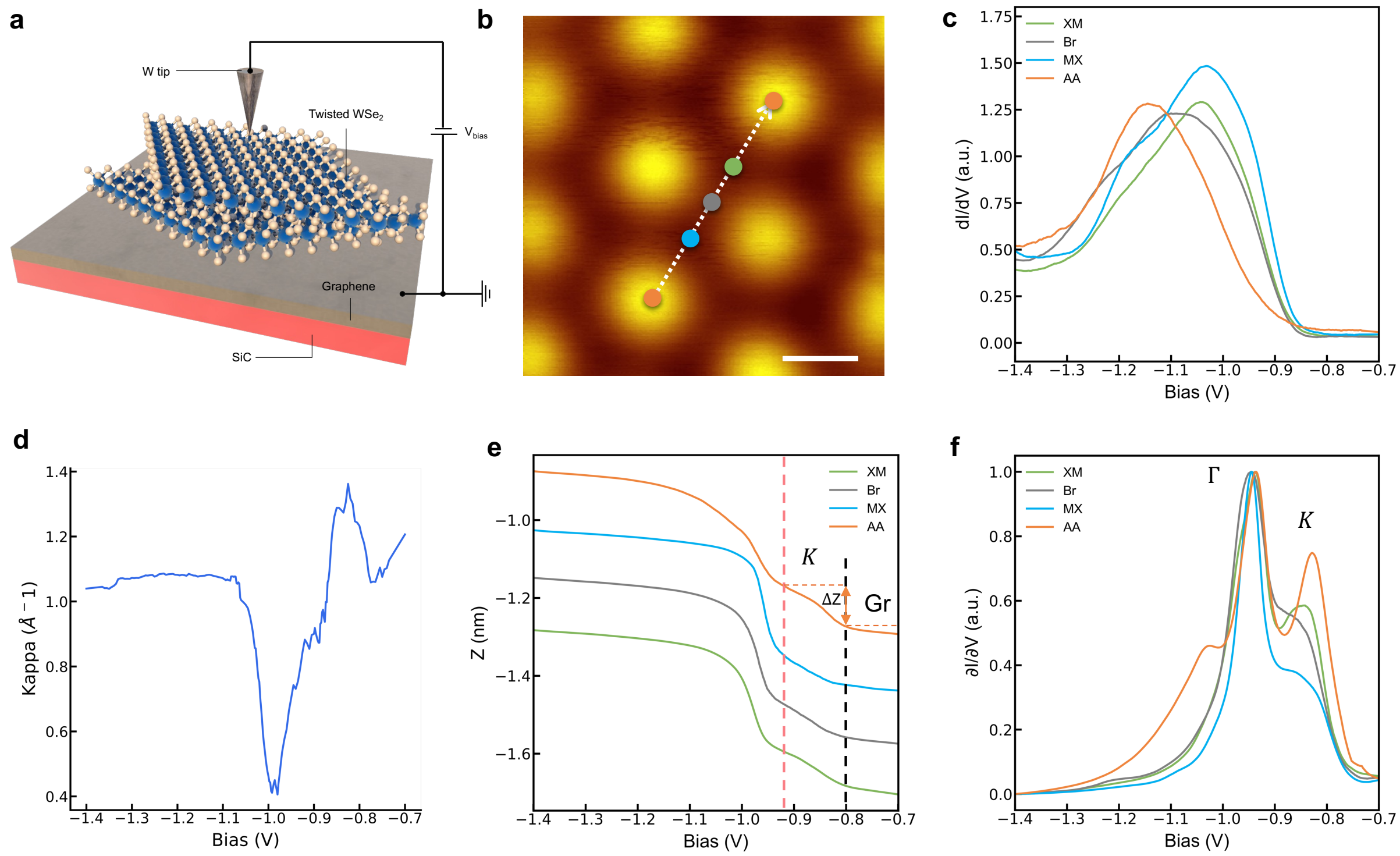

**Figure 2 | Γ and K valley moiré modulated tunneling spectroscopies. a,** The schematic STM experimental measurement set-up of a small angle tWSe$_2$ homobilayer. **b,** STM image showing the moiré pattern with a periodicity of 3.71 nm, which corresponds to a twist angle 5.1 degree. Scale bar: 2 nm. ($V_S = -2$ V, $I = -50$ pA). **c,** Constant height STS at four sites: AA, MX, Br, XM. ($V_{int} = -2.2$ V, $I = -100$ pA and $V_{amp} = 30$ mV) **d,** Measured decay constant, κ, as a function of bias voltage at the AA site. ($I_{int} = -100$ pA) **e,** *Z–V* spectroscopies respectively from sites AA, MX, Br and XM. ($I_{int} = -100$ pA) The red dashed line marks the transition from Γ valley to K valley and the black dashed line marks the transition from K valley to the underneath graphene state. **f,** Constant current STS at four sites. ($V_{int} = -1.4$ V, $I = -100$ pA and $V_{amp} = 20$ mV).

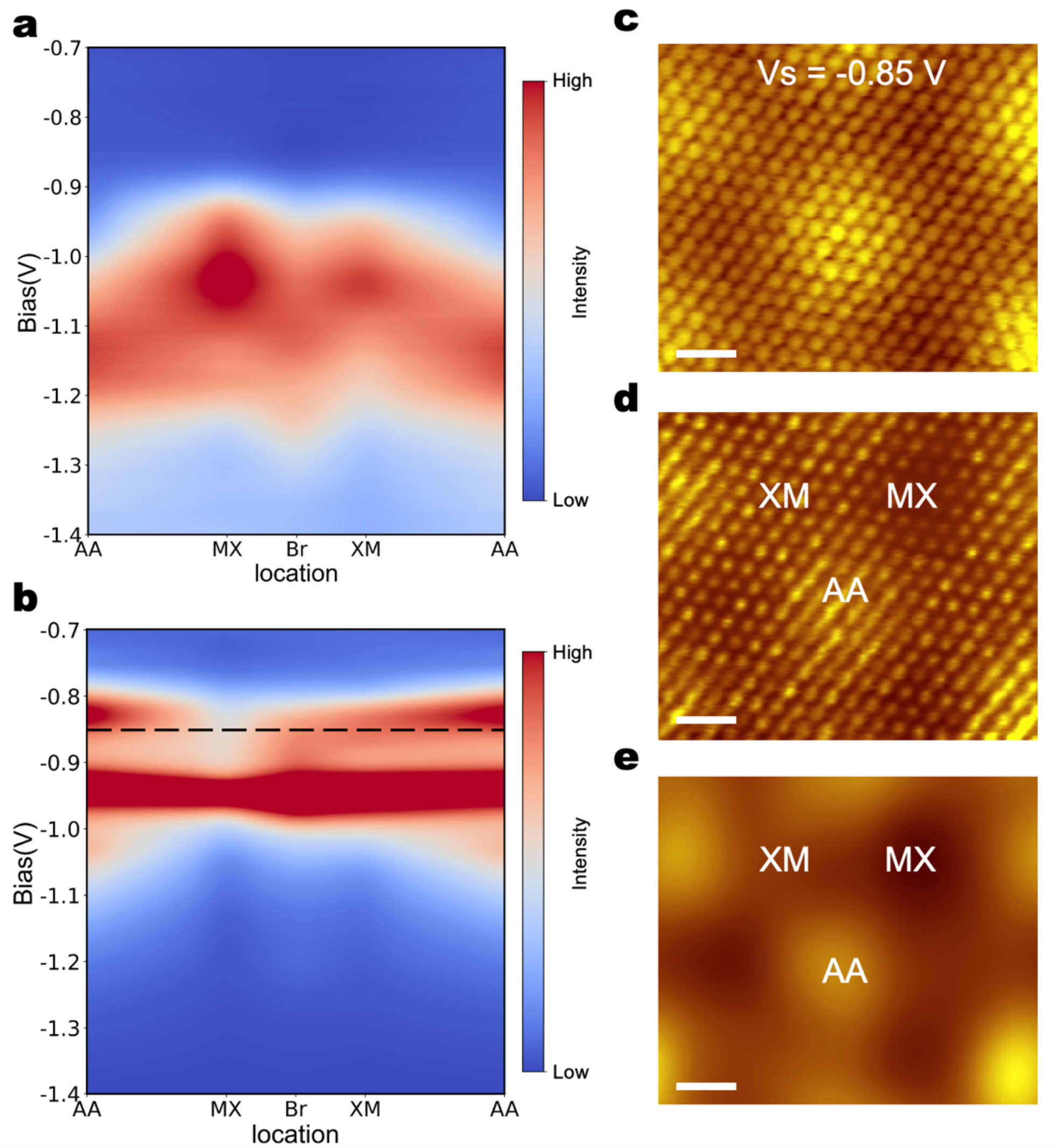

**Figure 3 | Real space mapping of layer pseudospin. a,** Constant-height and **b,** constant current STS color map along AA-MX-Br-XM-AA. ($V_{int} = -1.4$ V, $I = -100$ pA and $V_{amp} = 20$ mV). c-**e,** topography (c), constant height dI/dV (d), and filtered dI/dV mapping images (e) at $V_S = -0.85$ V, the voltage highlighted by the dashed line in b. Scale bar: 1nm. ($V_S = -0.85$ V, $I = -40$ pA, $V_{amp} = 20$ mV)

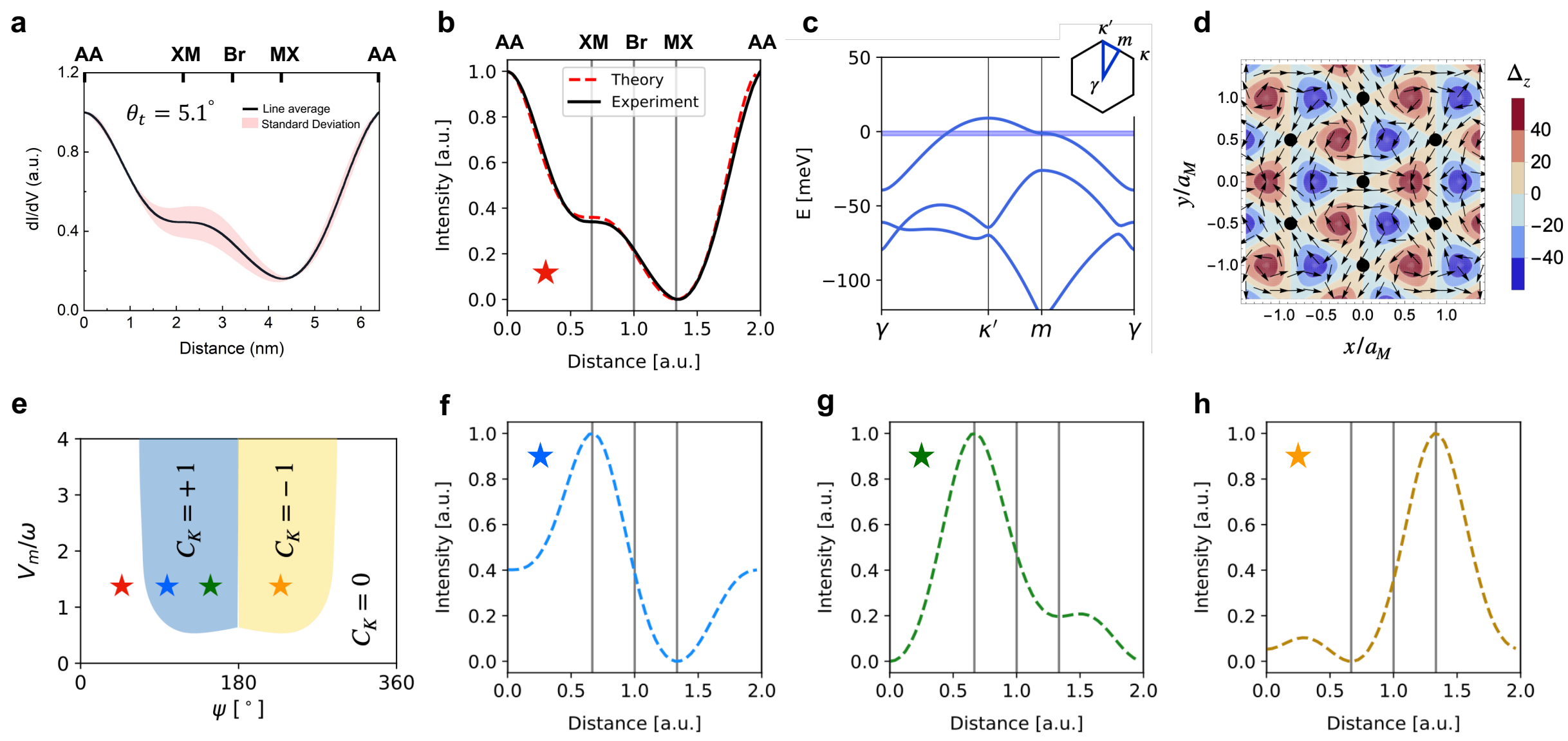

**Figure 4 | Obtaining continuum model parameters from STS. a,** averaged line profile and standard deviation of four different lines along the AA-XM-Br-MX-AA symmetric sites. For line profiles with uneven AA site intensities, they were symmetrized by multiplying position dependent linear coefficient and then renormalized with the intensity of AA site set as 1. **b,** Fit of the position-dependence of the layer-weighted LDOS near $K_{VBM}$. The black solid and red dashed lines are respectively experiment and theoretical fit. **c,** Moiré band-structure calculated with parameters obtained by fitting conductivity profile data along the high symmetry direction in Fig. 3e. The faint blue energy window marked where the theoretical LDOS curve in a is obtained. **d,** Layer-field $\boldsymbol{\Delta}$ with arrows corresponding to the x and y components and color-scale corresponding the z component. Black dots indicate moiré superlattice AA sites. **e,** Chern number of the topmost moiré band from the continuum model as a function of $\psi$ and $V_m/\omega$ at $\theta$=5.08°. The $C_K$=±1 regions are color coded. The red star labels the fitting showing in a. The blue, green and yellow star respectively correspond to blue, green and yellow curve in Fig. 4f-h in the continuum parameter space. **f-h,** Position-dependence of the layer-weighted LDOS near $K_{VBM}$ calculated with different set of parameters. Blue curve: $V_m = 13.6$ meV, $\psi = 90$ degrees and $\omega = 10$ meV. Green curve: same $V_m$ and $\omega$, $\psi = 150$ degrees; Yellow curve: $\psi = 230$ degrees.